\documentclass[preprint,12pt]{elsarticle}

\usepackage{amssymb,amsmath,amsfonts,amsthm}
\usepackage{color}

\newcommand{\bracket}[1]{\langle #1 \rangle}

\journal{Physica A}

\begin{document}

\begin{frontmatter}



\title{A stochastic model of the tweet diffusion \\ on the Twitter network}


\author{Tatsuro Kawamoto}

\address{Department of Physics, The University of Tokyo, Komaba, Meguro, Tokyo 153-8505, Japan}

\begin{abstract}
We introduce a stochastic model which describes diffusions of \textit{tweets} on the Twitter network.
By dividing the followers into generations, we describe the dynamics of the tweet diffusion as a random multiplicative process. 
We confirm our model by directly observing the statistics of the multiplicative factors in the Twitter data.
\end{abstract}


\end{frontmatter}



\section{Introduction}\label{Intro}
As a popular microblogging web service, a significant attention is paid to Twitter. 
One of the important points for users of Twitter is how well one's writings, or \textit{tweets}, are spreading through the network. 
Its significance is obvious from the facts that the counters which measure the popularity of tweets are everywhere on the web and that many companies are using Twitter as an advertising tool. 
There have already been many data related to Twitter \cite{Java07,Kwak10,Krishnamurthy08,Yang10,Cha10,Suh10,Galuba10} presented in the literature.

On Twitter, users can \textit{follow} other users and read their tweets without any approvals, thereby constructing a directed network among them.  
Users can also propagate tweets to their followers thanks to a characteristic function called \textit{retweet} \cite{FormalInformalRT}, which results in information diffusion. 
The aim of the present paper is to introduce a stochastic model for the tweet diffusion of the \textit{daily tweets} on the Twitter network. 
The reason why we investigate the daily tweets is because we can expect a universal behavior; 
in the case where we selected tweets with specific keywords, we might observe irregular behaviors depending on the characteristics of the keywords. 
As we will see, any tweet diffusion of the daily tweets seems to have a random multiplicative process as the underlying mechanism.

\begin{figure}[t]
\begin{center}
\includegraphics[width=0.6 \textwidth]{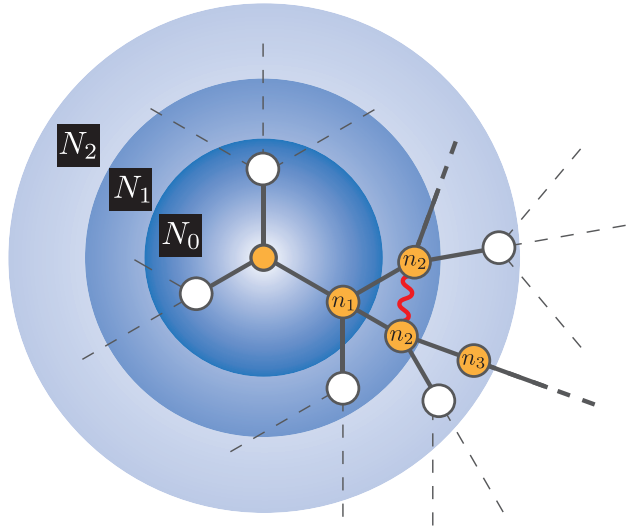}
\end{center}
\caption{
(Color online) 
Diffusion network on Twitter.
The node at the center represents the seed account and the linked nodes are the followers.
A solid line means that the tweet has diffused through the link by the retweet.
In the model analysis, we ignore the over-counting of followers such as the one illustrated by the wavy line, 
while we take account of it in the data analysis. 
}
\label{TwitterNetwork}
\end{figure}

We believe that the information diffusion on Twitter is different from some other web contents which are discussed in the literature. 
It has been known that the total spread of the information of many web contents obey lognormal distributions \cite{WuHuberman07,Wilkinson08,Yan11}. 
There, one can think of a discrete-time random multiplicative process; 
the additional number of spreads in one time step depends on the total number of spreads by the previous time steps and a decaying function of time, 
\textit{i.e.}, $N_{t+1} = N_{t}(1 + r_{t}X_{t})$, where $t$ is the discrete time, $N_{t}$ is the total number of spreads by time $t$, $X_{t}$ is a positive stochastic variable, and $r_{t}$ is the decaying function. 
It is a natural modeling for featured contents and contents that people search for. 
In the case of Twitter, however, it is unnatural to assume such a mechanism because we expect that tweets diffuse through the followers and the probability of the retweet activity should not depend on the total spread; 
users usually do not look for the tweets to retweet. 
They retweet because they receive the tweets.

In the present paper, we will propose to classify the followers into \textit{generations} depending on the distance from the seed account and consider a stochastic process along them. 
The diffusion process which we present would probably occur in many other networks, especially on the web. 
An advantage of researching the Twitter data is that we can confirm our model directly thanks to the Twitter API \cite{API}, which provides rich information.

The present paper is organized as follows. 
In Sec.~\ref{Model}, we will introduce a stochastic model of the tweet diffusion along the generations. 
In Sec.~\ref{DataAnalysis}, we will confirm that our model is indeed plausible by directly observing the stochastic variables of the model in the Twitter data. 
Note that we will fix a seed account when we analyze the data and there are some restrictions for the selection of the seed account (see Sec.~\ref{Restrictions}). 
Finally, after summarizing the present paper, 
we will argue what can be further expected in Sec.~\ref{conclusion}.

\section{Model}\label{Model}
Before we construct a model, let us first explain how a tweet diffuses by retweet on Twitter in detail.
Figure \ref{TwitterNetwork} shows a schematic picture of the tweet diffusion process.
Whenever a user generates a tweet, it will be sent to $N_{0}$ followers of the tweet owner, whom we call users in the zeroth generation.
Next, when $n_{1}$ users out of $N_{0}$ followers retweet, the original tweet will be sent to the followers of the $n_{1}$ retweeters; we call them users in the first generation.  
We label the number of the receivers in the first generation as $N_{1}$.
Such a chain of diffusion of a tweet continues until people stop retweeting or all the followers in the last generation are the users who already received the tweet \cite{Restrictions}. 
We will refer to the total number of receivers as $N_{\mathrm{tot}} = \sum_{g=0}^{\infty} N_{g}$ 
and the total retweet count as $n_{\textrm{\tiny RT}} = \sum_{g=1}^{\infty} n_{g}$, where $g$ stands for the label of the generation. 
While $N_{0}$ is simply the number of the followers of the seed account, 
$N_{g}$ for $g\ge1$ reads 
\begin{align}
N_{g} &= \sum_{f=1}^{n_{g}} k_{f} - c_{g}, \label{Ng} 
\end{align}
where $f$ stands for the label of each retweeter and $k_{f}$ stands for the number of his or her followers. 
The factor $c_{g}$ is the number of over-counting of the followers (\textit{e.g.}, the wavy line in Fig.~\ref{TwitterNetwork}). 
In the case where the network is close to the tree structure and the distribution of the number of followers is homogeneous, we can employ the approximation 
\begin{align}
N_{g} &\simeq n_{g} \sum_{k=0}^{\infty} k \, p_{g}(k) =: n_{g} \overline{k}_{g}, \label{NgApprox} 
\end{align}
where $k$ and $p_{g}(k)$ are the number of the followers of the retweeters in the $(g - 1)$th
generation and its distribution, respectively. 

Let us next estimate the number of the retweeters, $n_{g}$. 
Since there are $N_{g-1}$ candidates to generate the retweeters in the $g$th generation, we assume 
\begin{align}
&n_{g} = \beta_{g} N_{g-1},
\label{ng}
\end{align}
where $\beta_{g}$ is a variable which we call the retweet rate. 
Although $\beta_{g}$ is a discrete variable because $n_{g}$ and $N_{g-1}$ are integers, we treat it as if it were a continuous variable.
In Sec.~\ref{DataAnalysis}, we will observe that the retweet rate has a distribution over many incidents of tweet diffusion. 
We therefore regard $\beta_{g}$ as a continuous stochastic variable hereafter.

Combining eqs.~(\ref{NgApprox}) and (\ref{ng}), we have 
\begin{align}
&N_{m} = J_{m} N_{m-1} = \cdots = \prod_{g=1}^{m} J_{g} N_{0}, \label{NgRMP} \\
&n_{m} = \beta_{m} N_{m-1} = \cdots = \beta_{m} \prod_{g=1}^{m-1} J_{g} N_{0}, \label{ngRMP}
\end{align}
where 
\begin{align}
&J_{g} = \beta_{g} \overline{k}_{g}, 
\end{align}
which is a stochastic variable because $\beta_{g}$ is a stochastic variable. 
Although the probability distribution of $J_{1}$ may strongly depend on the characteristics of the seed account, 
$J_{g}$ for $g \ge 2$ are expected to obey a common probability distribution.
Therefore, the number of viewers of the tweet in each generation, $N_{g}$, is expressed as a random multiplicative process because of the hierarchical structure of the followers.
Note that the present model is not a standard percolation model, which assigns a stochastic variable to each follower, but a stochastic process with respect to each generation.
We do not consider the time dependence of the retweet rates since most of the tweets finish diffusing very quickly \cite{Kwak10}. 

In the following section, we will directly observe the statistics of the retweet rates $\beta_{g}$ and confirm that our modeling is indeed plausible.

\begin{figure}[t]
\begin{minipage}{0.5\hsize}
\begin{center}
\includegraphics[width=\textwidth]{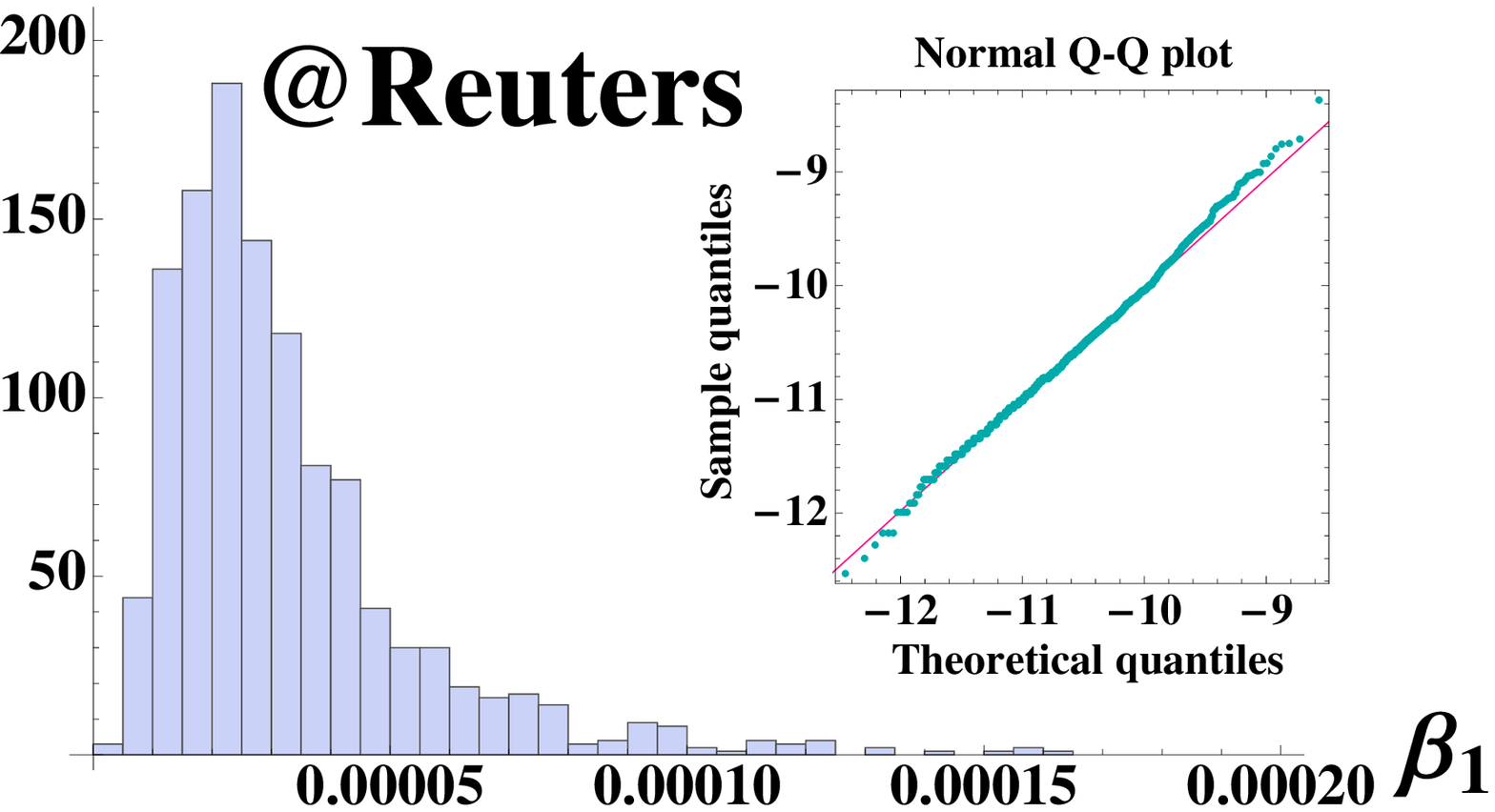}
(a)
\end{center}
\end{minipage}
\begin{minipage}{0.5\hsize}
\begin{center}
\includegraphics[width=\textwidth]{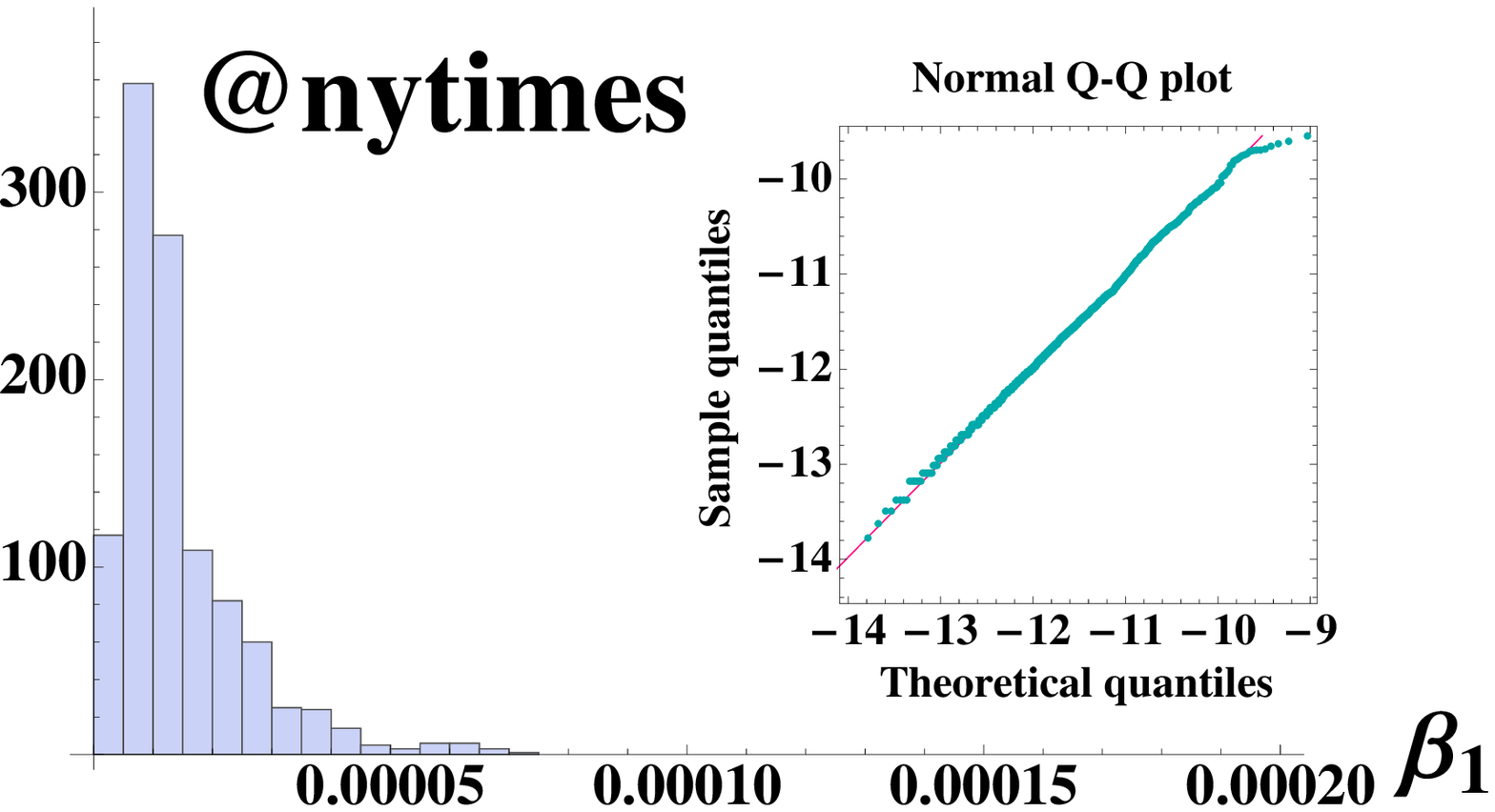}
(b)
\end{center}
\end{minipage}

\begin{minipage}{0.5\hsize}
\begin{center}
\includegraphics[width=\textwidth]{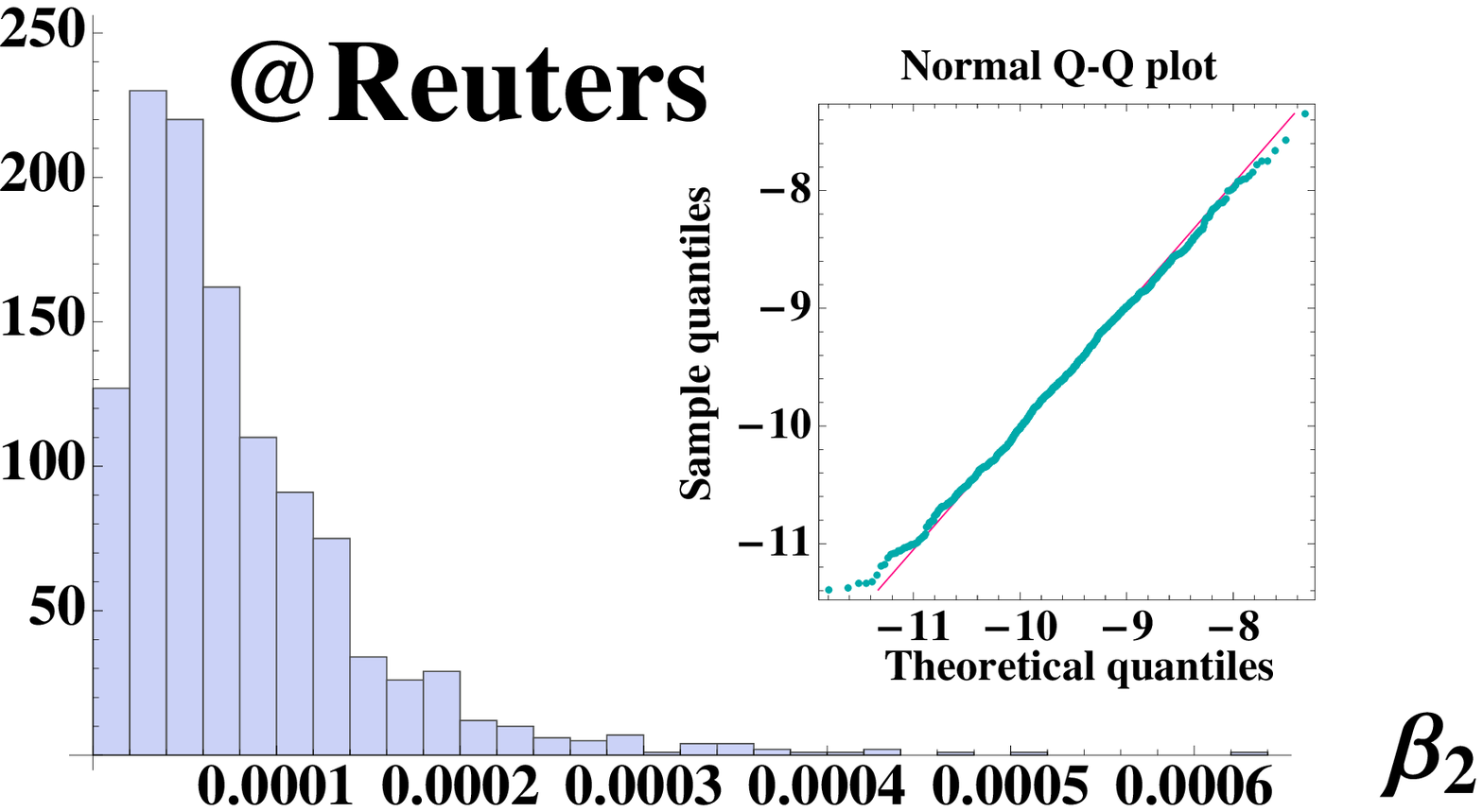}
(c)
\end{center}
\end{minipage}
\begin{minipage}{0.5\hsize}
\begin{center}
\includegraphics[width=\textwidth]{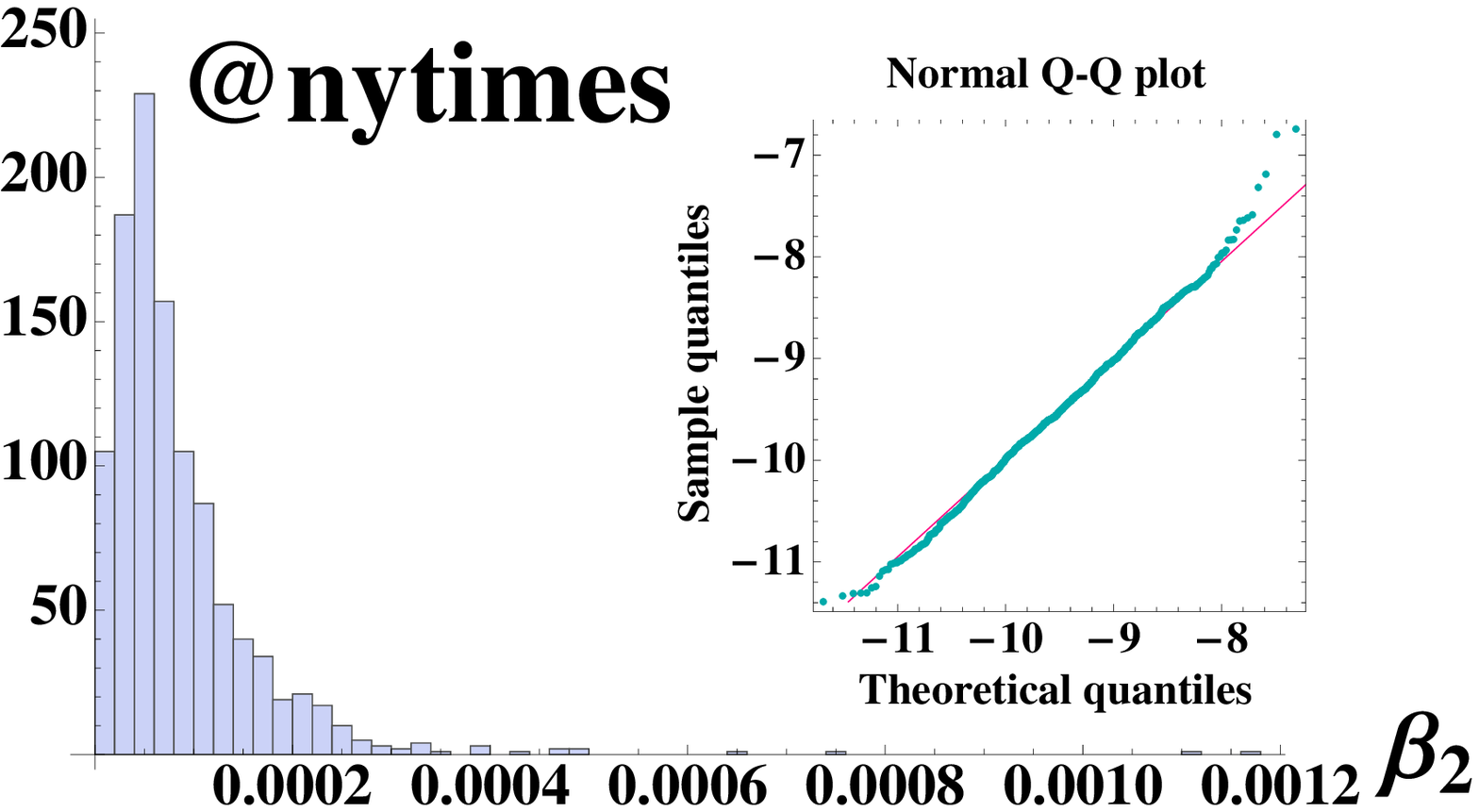}
(d)
\end{center}
\end{minipage}

\caption{
(Color online)
The histograms of the retweet rates and the normal Q-Q plots of the logarithms of the retweet rates. 
(a) and (c) are of $\beta_{1}$ and $\beta_{2}$ for @Reuters. (b) and (d) are of $\beta_{1}$ and $\beta_{2}$ for @nytimes. 
}
\label{betaX}
\end{figure}

\section{Data analysis for the retweet rate $\beta_{g}$}\label{DataAnalysis}
Using the data sampled by the tool Twitter API \cite{API}, 
we directly observed the behaviors of $\beta_{1}$ and $\beta_{2}$. 
We chose The New York Times (@nytimes) and Reuters Top News (@Reuters) for the seed accounts and 
sampled the diffusion data with $n_{2} > 0$. 
The data are summarized in Table \ref{DataTable}. 

\subsection{Possible errors, selection of the seed accounts, and restrictions}\label{Restrictions}
There are some inevitable errors in our data. 
We cannot sample the data of private users and there might be some miscounts in $n_{g}$ because the follow-followed relation might have changed by the time we sampled the data. 
In order to sample the data as accurately as possible, we need to select the seed accounts carefully; 
we chose the seed accounts which tweet frequently and the number of whose followers are not changing rapidly so that we can expect the network around the seed account is almost static during the period of sampling. 
In order to see the statistical behavior clearly, it is good to choose an account with a large number of followers and high retweet rates \cite{RTlimitation}. 
When we analyze the retweet rate $\beta_{g}$, we take into account the factor of over-counting $c_{g}$ in Eq.~(\ref{Ng}), and thus we do not assume a tree structure nor the homogeneity of the distribution of the followers.

\subsection{Result}
Figures \ref{betaX}(a) and \ref{betaX}(b) show the histograms of $\beta_{1}$ and their normal Q-Q plots \cite{QQplot}. 
They show that the retweet rate $\beta_{1}$ seems to obey lognormal distributions with slight additive shifts, \textit{i.e.} 
\begin{align}
\beta_{1} =  \mathrm{e}^{ \omega_{1} } + \delta_{1}, 
\end{align}
where $\omega_{1}$ obeys Gaussian distributions $\mathcal{N}(\mu_{1}, \sigma^{2}_{1})$ with $\mu_{1}$ being the mean and $\sigma^{2}_{1}$ being the variance of $g=1$. 
For $\beta_{1}$, the mean $\mu_{1}$ and the variance $\sigma^{2}_{1}$ seem to depend strongly on the character of the seed account. 
The slight additive shift might be due to the systematic activities by Twitter bots.

We expect that the retweet rate $\beta_{2}$ also obeys lognormal distributions with slight additive shifts. 
Figures \ref{betaX}(c) and \ref{betaX}(d) show the histograms of $\beta_{2}$ and their normal Q-Q plots; 
they indeed indicate the lognormal behaviors. 
For $\beta_{2}$, the mean $\mu_{2}$ and the variance $\sigma^{2}_{2}$ are very close for both of the seed accounts; 
it seems to be plausible to model that the retweet rate $\beta_{g}$ obeys a common probability distribution for $g \ge 2$. 

In Table \ref{DataTable}, we listed the averages of the over-counting of the users in the first generation, \textit{i.e.} $c_{1} / \sum_{f=1}^{n_{1}} k_{f}$ in Eq.~(\ref{Ng}). 
The over-counting of users are less than $5\%$ on average, and thus the networks around the seed accounts have almost the tree structures. 
Although it is still doubtful whether the tree-structure approximation is appropriate in all generations, 
it is hard to imagine a drastic qualitative change to the diffusion phenomenon due to the loop correction since there is no back flow.

\begin{figure}[t]
\begin{minipage}{0.5\hsize}
\begin{center}
\includegraphics[width=\textwidth]{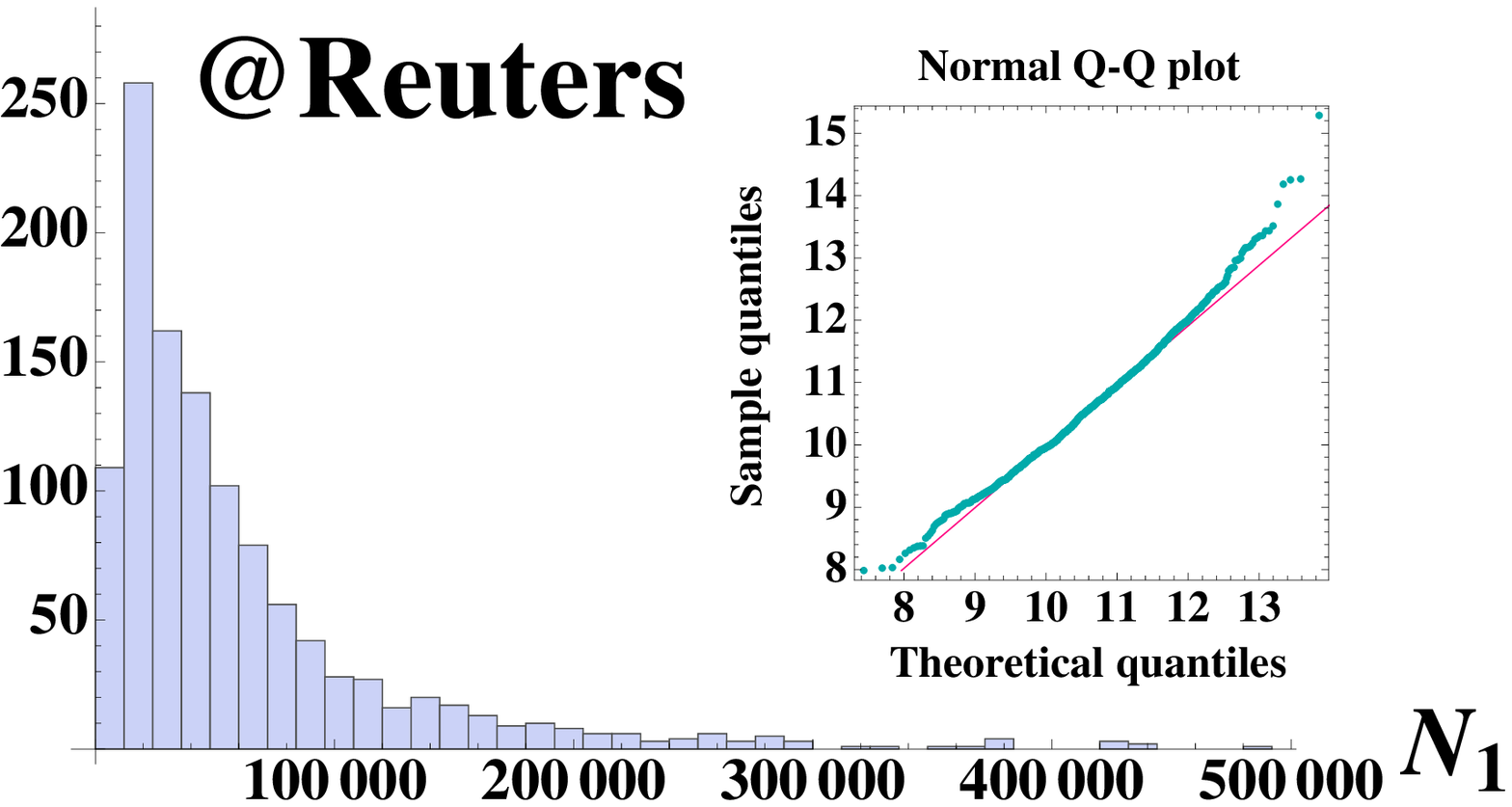}
(a)
\end{center}
\end{minipage}
\begin{minipage}{0.5\hsize}
\begin{center}
\includegraphics[width=\textwidth]{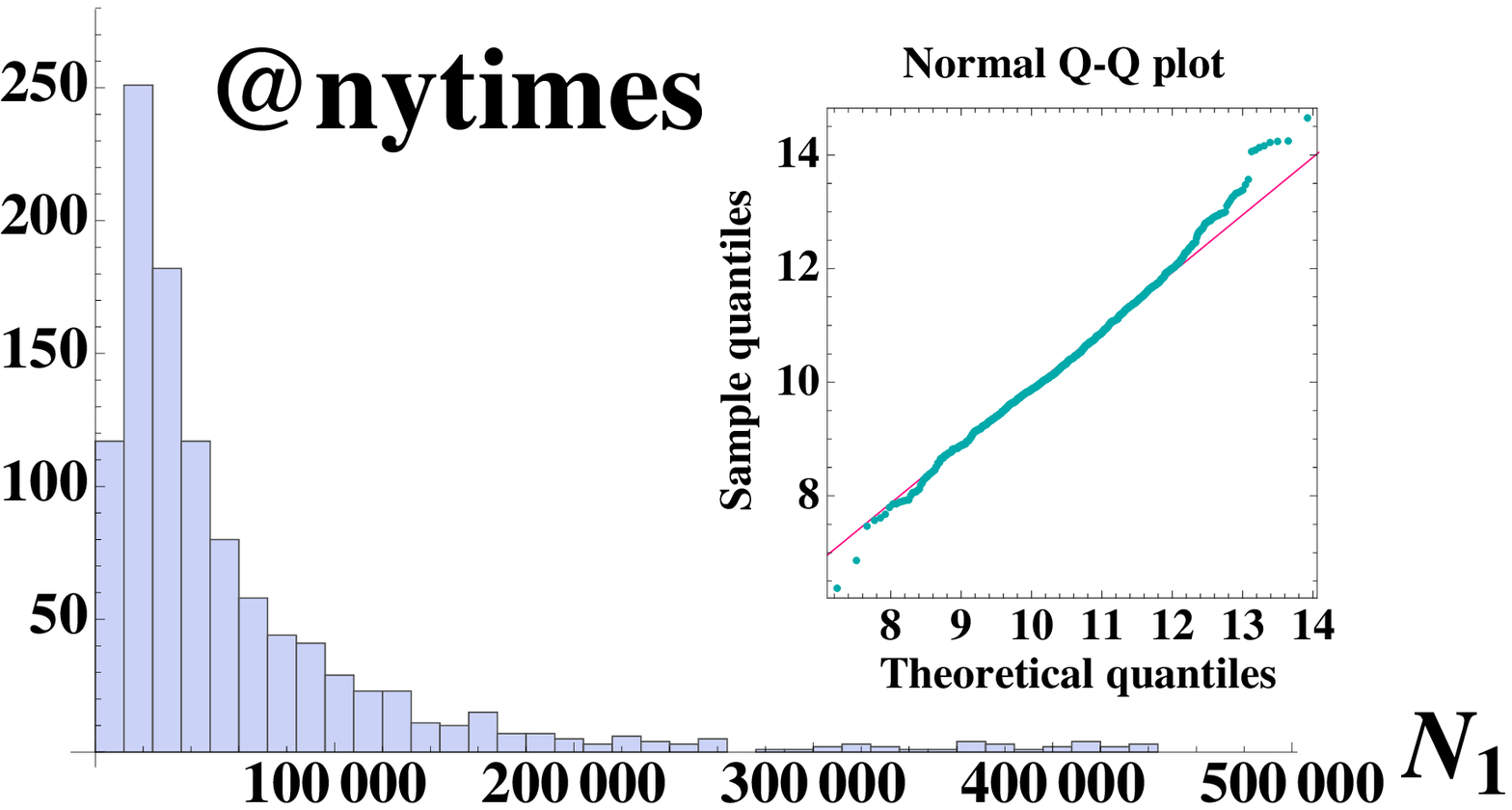}
(b)
\end{center}
\end{minipage}

\begin{minipage}{0.5\hsize}
\begin{center}
\includegraphics[width=\textwidth]{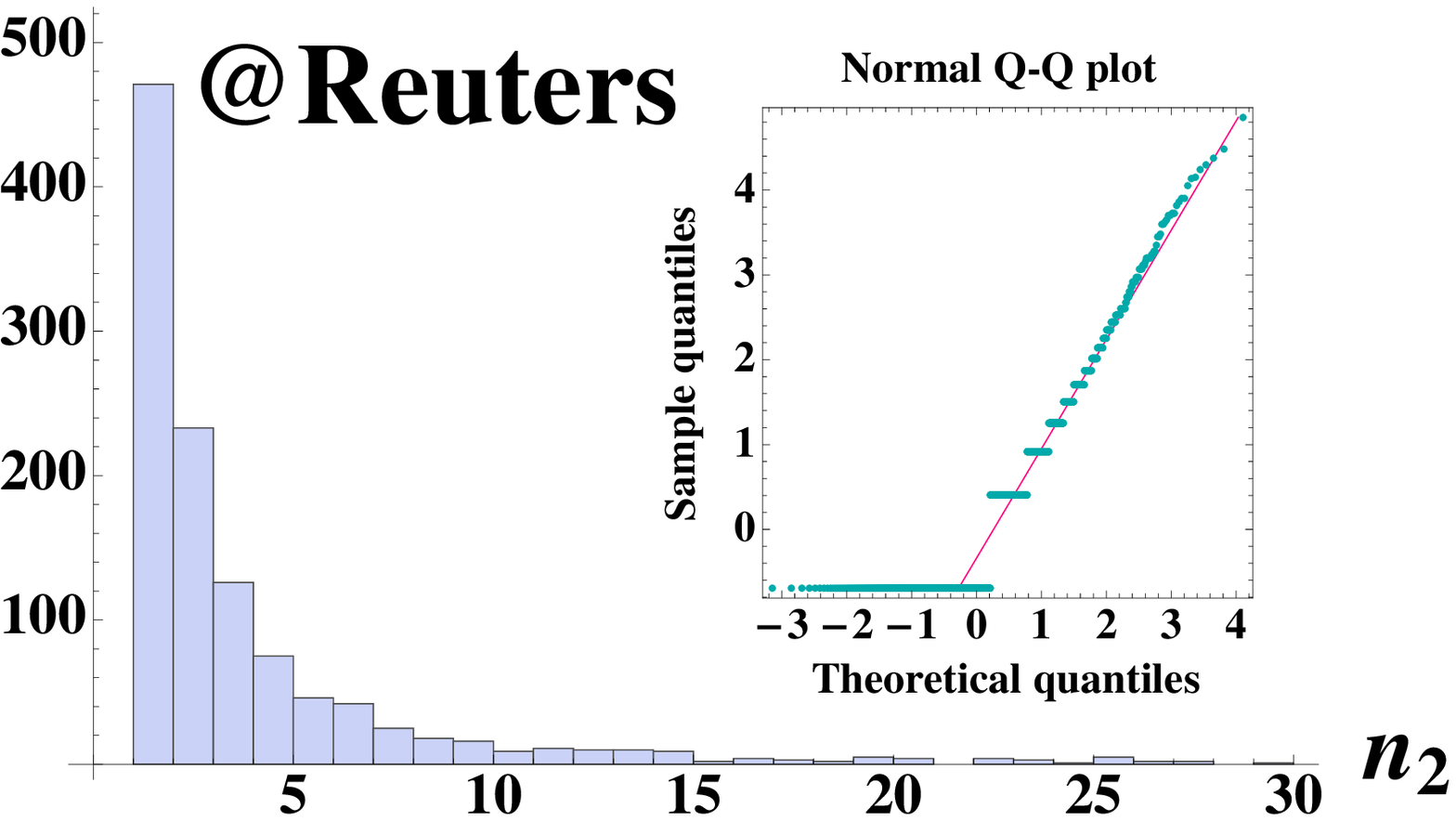}
(c)
\end{center}
\end{minipage}
\begin{minipage}{0.5\hsize}
\begin{center}
\includegraphics[width=\textwidth]{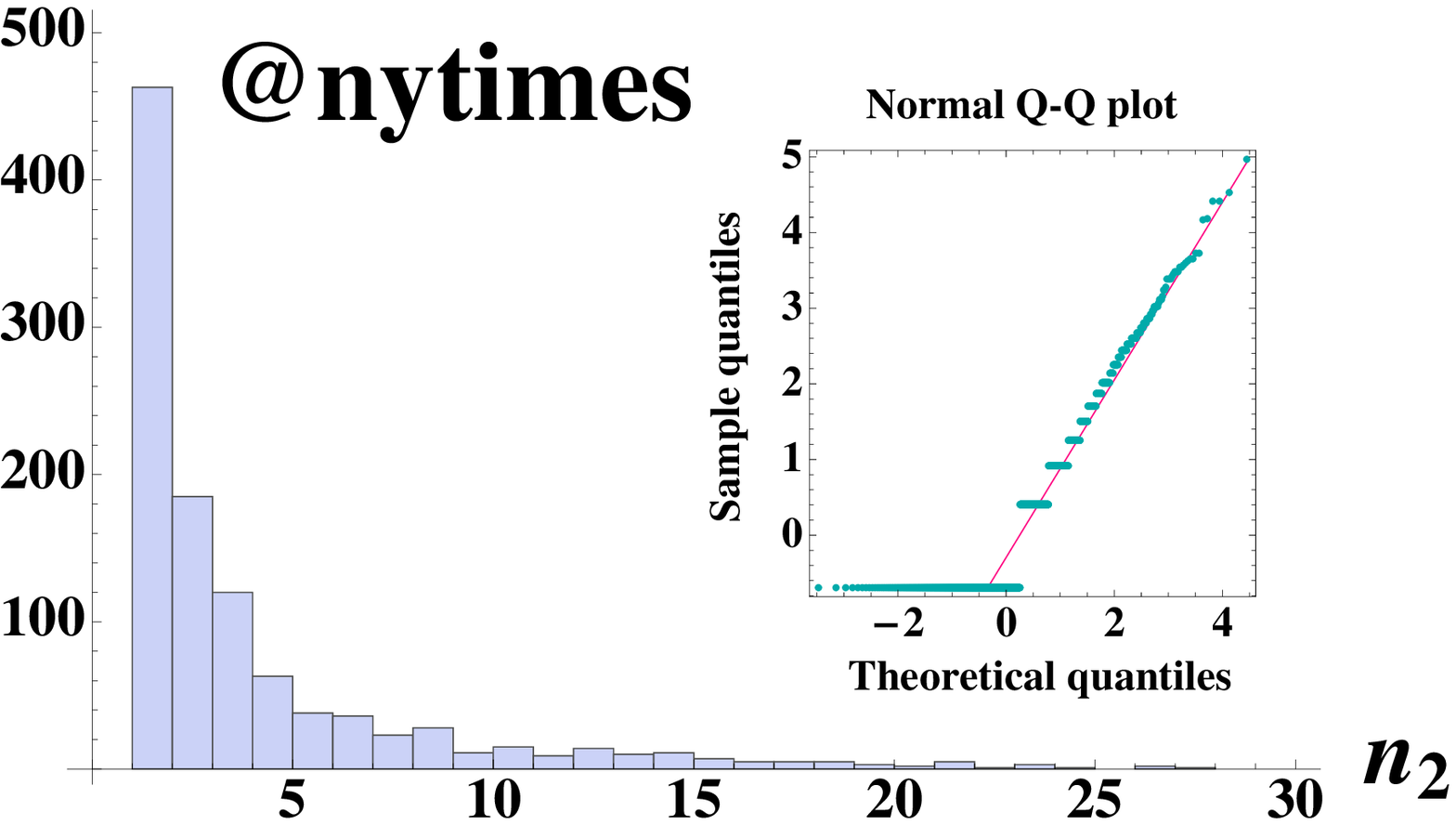}
(d)
\end{center}
\end{minipage}

\caption{
(Color online)
The histograms of the retweet rates and the normal Q-Q plots of the logarithms of the number of followers and the retweet count in the second generation. 
(a) and (c) are of $N_{1}$ and $n_{2}$ for @Reuters. (b) and (d) are of $N_{1}$ and $n_{2}$ for @nytimes. 
}
\label{N1n2}
\end{figure}

Since we are fixing the seed account, $N_{0}$ is a constant and the distribution of $\beta_{1}$ is proportional to that of $n_{1}$. 
The number of followers in the first generation, $N_{1}$, and the number of retweeters among them, $n_{2}$, can take different values for each sample. 
As are shown in Figs.~\ref{N1n2}(c) and \ref{N1n2}(d), both of them obey lognormal distributions and they are not independent of each other. 
The correlation coefficients of $n_{2}$ and $N_{1}$, $\rho(n_{2}, N_{1})$, have large positive values (see Table \ref{DataTable}); 
the correlation coefficient varies from $-1$ to $1$ and is calculated by 
\begin{align}
\rho(n_{2}, N_{1}) = \frac{ \bracket{n_{2} N_{1}} - \bracket{n_{2}} \bracket{N_{1}} }{ \sqrt{ \bracket{ n_{2}^{2} } - \bracket{ n_{2} }^{2} } \sqrt{ \bracket{ N_{1}^{2} } - \bracket{ N_{1} }^{2} } }.
\end{align}
Our result that the retweet rate $\beta_{2}$ obeys a lognormal distribution is plausible because lognormal distributions have the reproductive property, \textit{i.e.} for 
two stochastic variables $X_{1}$ and $X_{2}$ which obey lognormal distributions, 
\begin{align}
p(\ln X_{1}) * p(\ln X_{2})
&= \mathcal{N}(\mu_{1}, \sigma^{2}_{1}) * \mathcal{N}(\mu_{2}, \sigma^{2}_{2}) 
= \mathcal{N}(\mu_{1}+\mu_{2}, \sigma^{2}_{1}+\sigma^{2}_{2}), 
\end{align}
where $*$ stands for the convolution.

\begin{table}
\small
\centering
\begin{tabular}{ccc}
&@Reuters & @nytimes \\
\hline
Number of seed tweets & $1352$ & $1140$\\
\hline
Period & from $\textrm{Jun. }26,\,2012$ & from $\textrm{Jun. }19,\,2012$\\
	& to $\textrm{Aug. }9,\,2012$ & to $\textrm{Aug. }18,\,2012$\\
\hline
$N_{0}$ & $1\,940\,477$ & $5\,882\,680$ \\
\hline
$\bracket{ n_{\mathrm{RT}} }$ & $74.0$ & $97.3$\\
\hline
$(\bracket{\beta_{1}}, \,\bracket{\beta_{2}})$ & $(3.27\times 10^{-5}, \,7.90 \times 10^{-5})$ & $(1.43\times 10^{-5}, \,8.52 \times 10^{-5})$\\
\hline
$(\bracket{N_{1}}, \,\bracket{n_{2}} )$ & $(75\,276, \,4.46)$ & $(76\,533, \,4.54)$\\
\hline
$\rho(n_{2}, \,N_{1})$ & $0.468$ & $0.481$\\
\hline
$\bracket{c_{1} / \sum_{f=1}^{n_{1}} k_{f}}$ & $0.0471$ & $0.0470$\\
\hline
Fitting parameters & $\mu_{1} = -10.51$ & $\mu_{1} = -11.51$\\
for $\beta_{1} = \mathrm{e}^{\omega_{1}} + \delta_{1}$	& $\sigma^{2}_{1} = 0.6$ & $\sigma^{2}_{1} = 0.77$\\
$p(\omega_{1}) = \mathcal{N}(\mu_{1}, \sigma^{2}_{1})$ 	& $\delta_{1} = 0$ & $\delta_{1} = 1.0 \times 10^{-6}$\\
\hline
Fitting parameters & $\mu_{2} = -9.65$ & $\mu_{2} = -9.51$\\
for $\beta_{2} = \mathrm{e}^{\omega_{2}} + \delta_{2}$	& $\sigma^{2}_{2} = 0.68$ & $\sigma^{2}_{2} = 0.69$\\
$p(\omega_{2}) = \mathcal{N}(\mu_{2}, \sigma^{2}_{2})$ 	& $\delta_{2} = -1.0\times 10^{-5}$ & $\delta_{2} = -1.0 \times 10^{-5}$\\
\hline
Fitting parameters & $\mu = 10.64$ & $\mu = 10.58$\\
for $N_{1} = \mathrm{e}^{\omega} + \delta$	& $\sigma^{2} = 0.99$ & $\sigma^{2} = 1.04$\\
$p(\omega) = \mathcal{N}(\mu, \sigma^{2})$ 	& $\delta = 0$ & $\delta = 2000$\\
\hline
Fitting parameters & $\mu = 0.48$ & $\mu = 0.49$\\
for $n_{2} = \mathrm{e}^{\omega} + \delta$	& $\sigma^{2} = 1.12$ & $\sigma^{2} = 1.23$\\
$p(\omega) = \mathcal{N}(\mu, \sigma^{2})$	& $\delta = 0.5$ & $\delta = 0.5$\\
\hline
\end{tabular}
\caption{ Data of tweet diffusions from @Reuters and @nytimes. 
The angular bracket $\bracket{\cdots}$ stands for the sample average.}
\label{DataTable}
\end{table}

\section{Model analysis: estimation of the diffusion range}
From the model which we introduced above, we can estimate how much of the retweet rate $\beta^{\mathrm{th}}(m)$ is required to reach the $m$th generation on average and 
the average of the total number of retweets, $\bracket{n_{\mathrm{RT}}}$, for given parameters. 
In this section, we restrict ourselves to the case where the retweet rates $\beta_{g}$ are independent of each other and their averages have a common value $\bracket{\beta}$. 

\begin{figure}[t]
\begin{minipage}{0.5\hsize}
\begin{center}
\includegraphics[width=\textwidth]{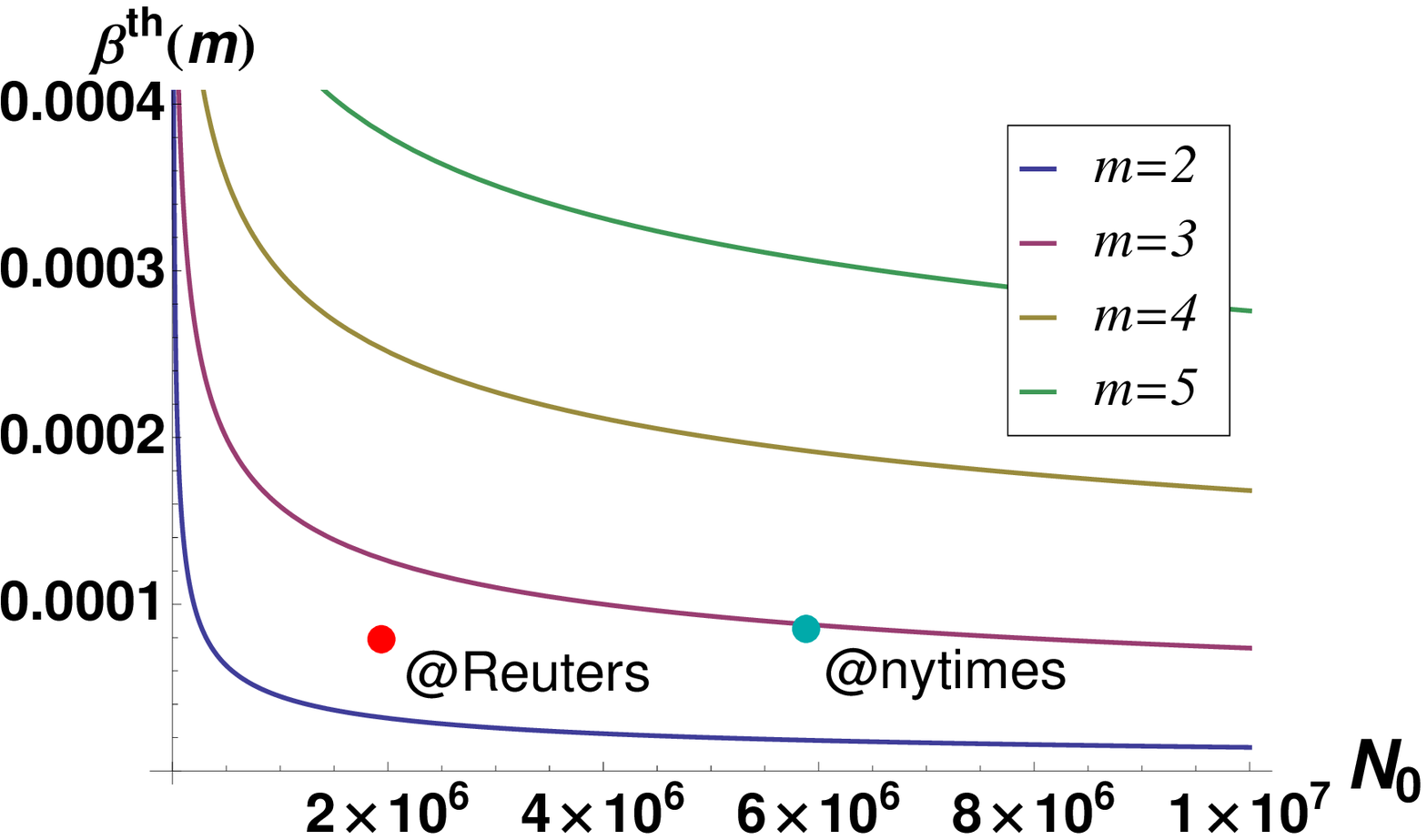}
(a)
\end{center}
\end{minipage}
\begin{minipage}{0.5\hsize}
\begin{center}
\includegraphics[width=\textwidth]{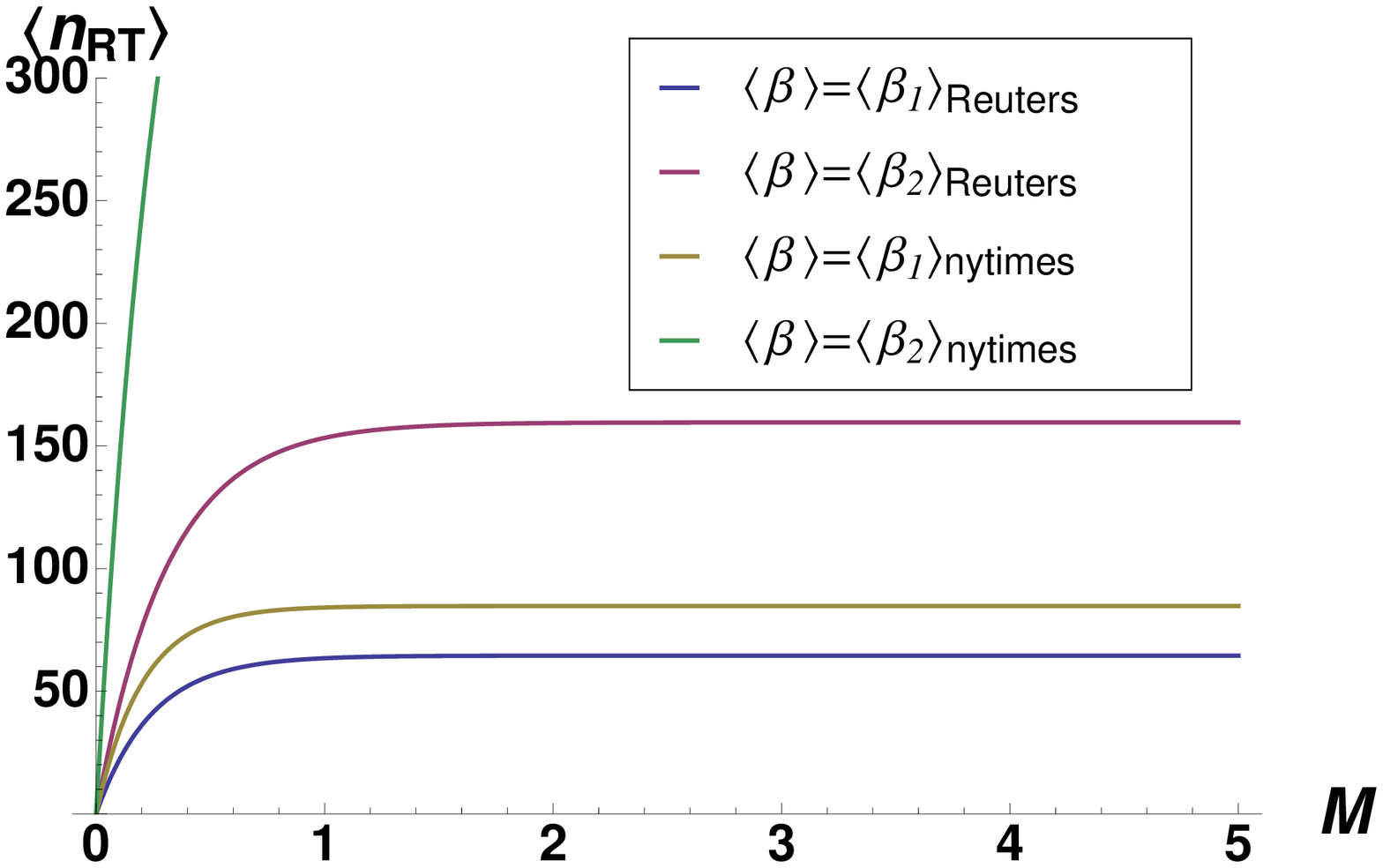}
(b)
\end{center}
\end{minipage}

\caption{
(Color online)
(a) Threshold where the diffusion reaches the $m$th generation on average. We set $\overline{k} = 500$. The points are for @Reuters and @nytimes in the case where we assumed $\bracket{\beta} = \bracket{\beta_{2}}$.
(b) The average of the total number of retweets $\bracket{n_{\mathrm{RT}}}$ as a function of the diffusion range $M$ for various values of the average retweet rate $\bracket{\beta}$.  
We set $\overline{k} = 500$ and plotted the cases where 
$\bracket{\beta}$ equals $\bracket{\beta_{g}}$ of @Reuters and @nytimes. 
We set the values of them for $N_{0}$, respectively. 
}
\label{ModelAnalysisPlot}
\end{figure}

According to Eq.~(\ref{ngRMP}), the average of the number of retweets in the $m$th generation reads 
$\bracket{n_{m}} = N_{0} \overline{k}^{m-1} \bracket{\beta}^{m}$. 
Then we have the threshold for the retweet rate where the diffusion reaches the $m$th generation on average, 
\textit{i.e.} $\bracket{n_{m}} \ge 1$: 
\begin{align}
\beta^{\mathrm{th}}(m) = \left( N_{0} \overline{k}^{m-1} \right)^{-\frac{1}{m}} = N_{0}^{-\frac{1}{m}} \overline{k}^{\frac{1}{m}-1}.  \label{betathreshold}
\end{align}
The behavior of Eq.~(\ref{betathreshold}) is exemplified in Fig.~\ref{ModelAnalysisPlot}(a); 
in the case of the seed accounts which we investigated, the tweets diffuse up to the second or the third generation (see $\bracket{\beta_{1}}$ and $\bracket{\beta_{2}}$ in Table~\ref{DataTable}).
While we employed the mean value of $n_{m}$ in the definition of the threshold, 
it is also plausible to consider the median of $n_{m}$ instead. 

For a given range $M$ of the diffusion, it is straightforward to calculate the average of the total number of retweets, 
\begin{align}
\bracket{ n_{\mathrm{RT}} } &= \sum_{g=1}^{M} \bracket{ n_{g} } = N_{0} \bracket{\beta} \sum_{g=0}^{M-1} \left( \bracket{\beta} \overline{k} \right)^{g}
= N_{0} \bracket{\beta} \frac{1 - \left( \bracket{\beta}\overline{k} \right)^{M}}{1 - \bracket{\beta}\overline{k}}. \label{nRTestimation}
\end{align}
The behavior of Eq.~(\ref{nRTestimation}) is exemplified in Fig.~\ref{ModelAnalysisPlot}(b); 
it shows that $\bracket{ n_{\mathrm{RT}} }$ is not very sensitive to the diffusion range $M$ 
in the case where $\bracket{\beta}\overline{k}$ is small.

\section{Conclusion and Discussion}\label{conclusion}
We decomposed the diffusion of the daily tweets into the dynamics along the generations of the followers; 
the dynamics of retweet activities can be modeled as a random multiplicative process with respect to the generation.
We directly observed the multiplicative factors from the actual data of Twitter and confirmed that our model is indeed plausible. 
We found that the multiplicative factors roughly obey lognormal distributions. 
The important point about our model is that the diffusion occurs owing to the repetition of cooperative activities along the followers and thus, as far as we know, it belongs to a different class of information diffusion compared to the ones in the literature \cite{WuHuberman07,Wilkinson08,Yan11}. 
We also believe that Twitter is not the only network where such a diffusion process occurs. 

The model of the present paper is a minimal model. 
In order to estimate the behavior of diffusion precisely, the approximation of the tree structure is obviously too rough; 
it is necessary to embed the information about the rate of over-counting and the heterogeneity of the network. 
We also neglected the correlation between the retweet rates. 
We will consider its effect in a future study which may be published elsewhere. 

For the accuracy of the data, we sample the data of the news accounts only in the present paper. 
If Twitter API were updated and we tried a different way of sampling the data, we would be able to analyze the behaviors of many other accounts. 
Then we can proceed to a more quantitative analysis; 
for example, we would be able to measure the range of diffusion for each diffusion process.

\section{Acknowledgments}
The author thanks Naomichi Hatano, Tomotaka Kuwahara, and Tomohiko Konno for many useful discussions and Chris Wiggins for valuable comments.
This work was partially supported by the Aihara Innovative Mathematical Modelling Project.





\bibliographystyle{elsarticle-num}
\bibliography{<your-bib-database>}

\begin{thebibliography}{00}


\bibitem{Java07} A. Java, X. Song, T. Finin, and B. Tseng: 
Proc. of the 9th WebKDD and 1st
SNA-KDD 2007 workshop on Web mining and social network analysis (ACM) (2007) 56-65.

\bibitem{Kwak10} H. Kwak, C. Lee, H. Park, and S. Moon: 
Proc. of 19th Int. World Wide Web Conf. (WWW) (2010) 591-600.

\bibitem{Krishnamurthy08} B. Krishnamurthy, P. Gill, and M. Arlitt: 
Proc. of the 1st workshop on Online social networks (ACM) (2008) 19-24.

\bibitem{Yang10} J. Yang and S. Counts: 
Proc. of the Fourth International AAAI Conf. on Weblogs and Social Media (2010).

\bibitem{Cha10} M. Cha, H. Haddadi, F. Benevenuto, and K. P. Gummadi: 
Proc. of the Fourth Int. AAAI Conf. on Weblogs and Social Media (2010).

\bibitem{Suh10} B. Suh, L. Hong, P. Pirolli, and E. H. Chi: 
Proc. of the IEEE Second International Conference on Social Computing (SocialCom) (2010) 177-184.

\bibitem{Galuba10}{
W. Galuba, K. Aberer, D. Chakraborty, Z. Despotovic, and W. Kellerer: 
WOSN'10 Proc. of the 3rd Conf. on Online social networks (2010) 3.}

\bibitem{FormalInformalRT}{
Note that the word retweet is sometimes used for two different meanings in the literature.  
The retweet button was first introduced at the end of 2009. 
Retweeting used to be simply a name of custom on Twitter to transfer the tweet of another user; 
it is called \textit{informal retweet} nowadays, while the retweeting by clicking the retweet button is called \textit{formal retweet}. 
Galuba \textit{et al}. \cite{Galuba10} also introduced a model of tweet diffusion in a very different manner from ours, 
but they limited themselves to the URL-embedded tweets and counted the informal retweets, 
whereas we analyze tweets in general and count the formal retweets in the present paper.
}



\bibitem{WuHuberman07} F. Wu and B. A. Huberman: 
Proc Natl Acad Sci USA {\bf104}(45) (2007) 17599-17601. 


\bibitem{Wilkinson08} D. M. Wilkinson: 
Proc. of the 2008 ACM Conf. on E-Commerce (2008) 302-309.

\bibitem{Yan11} M. Yan and M. Gerstein, PLoS One {\bf 6} (2011) 5, e19917.



\bibitem{API} https://dev.twitter.com/docs/api, https://dev.twitter.com/docs/streaming-api

\bibitem{Restrictions} As long as the tweet owner is a public user, anyone can read the tweet and any user has the right to retweet, and therefore the chain of retweets among the followers is not the only way in which tweets diffuse. 
We do not treat such other processes in the present paper; 
we only counted the formal retweets \cite{FormalInformalRT} by non-private users 
because we believe that it gives the major contribution to the daily tweet diffusion. 



\bibitem{RTlimitation} We omitted the data with more than 800 retweets because Twitter API seems to fail to count the retweets correctly in such cases. 

\bibitem{QQplot} M. B. Wilk and R. Gnanadesikan, Biometrika \textbf{55} (1968) 1-17.

\end{thebibliography}



\end{document}